\documentclass[12pt]{article}
\usepackage{latexsym}
\usepackage{amsmath,,calrsfs}
\usepackage{amsfonts}
\usepackage{amssymb}
\usepackage{amscd}
\usepackage{bbm}
\usepackage{fancybox}
\usepackage{cite}
\usepackage{amsmath,amsfonts,amsbsy}
\usepackage{pstricks,pst-node}
\usepackage[small,bf,hang]{caption2}
\usepackage{graphicx}
\usepackage{epsfig}
\usepackage{psfrag}
\usepackage{comment}

\usepackage{float}

\psset{unit=1.3cm,linewidth=.5pt,radius=.2}  

\usepackage{multirow}                     
\usepackage{float}                          
\usepackage{lscape}                         
\usepackage{bm}

\newcommand{\bb}{\bar\beta}

\newcommand{\beq}{\begin{equation}}
\newcommand{\eeq}{\end{equation}}
\newcommand{\bi}{\begin{itemize}}
\newcommand{\ei}{\end{itemize}}
\newcommand{\bt}{\begin{tabular}}
\newcommand{\et}{\end{tabular}}
\newcommand{\bc}{\begin{center}}
\newcommand{\ec}{\end{center}}

\newcommand{\be}{\begin{equation}}
\newcommand{\ee}{\end{equation}}
\newcommand{\bea}{\begin{eqnarray}}
\newcommand{\eea}{\end{eqnarray}}
\newcommand{\ba}{\begin{array}}
\newcommand{\ea}{\end{array}}

\def\bbox{{\,\lower0.9pt\vbox{\hrule \hbox{\vrule height 0.2 cm
\hskip 0.2 cm \vrule height 0.2 cm}\hrule}\,}}
\newcommand{\dsl}{\pa \kern-0.5em /}

\font\mybb=msbm10 at 12pt
\def\bb#1{\hbox{\mybb#1}}

\def\bR {\bb{R}}

\def\bfo{\mbox{\boldmath $\omega$}}

\begin{document}

\begin{titlepage}
\begin{center}

\hfill  DAMTP-2017-7, ICCUB-17-04

\vskip 1.5cm

{\Large \bf  Tachyons in the Galilean limit}

\vskip 2cm

{\large {\bf  Carles Batlle\,${}^1$, Joaquim Gomis\,${}^2$, \\ 
\vskip 5pt 
Luca Mezincescu\,${}^3$ and Paul K.~Townsend\,${}^4$} }

\vskip 1cm

{\em $^1$ \hskip -.1truecm
\em Departament de Matem\`atiques and IOC, 
Universitat Polit\`ecnica de Catalunya \\
EPSEVG, Av. V. Balaguer 1, E-08808 Vilanova i la Geltr\'u, Spain \vskip 5pt }
{email: {\tt carles.batlle@upc.edu}}

\vskip 5pt

{\em $^2$ \hskip -.1truecm
\em  Departament de F\'{\i}sica Qu\`antica i Astrof\'{\i}sica\\ and Institut de Ci\`encies del Cosmos (ICCUB), 
Universitat de Barcelona, \\Mart\'{\i} i Franqu\`es 1, E-08028 Barcelona, Spain  \vskip 5pt }
{email: {\tt gomis@fqa.ub.edu}}

\vskip 5pt

{\em $^3$ \hskip -.1truecm
\em Department of Physics, University of Miami, \\
P.O. Box 248046, Coral Gables, \\ FL 33124, USA\vskip 5pt }

{email: {\tt mezincescu@physics.miami.edu}} 

\vskip 5pt

{\em $^4$ \hskip -.1truecm
\em  Department of Applied Mathematics and Theoretical Physics,\\ Centre for Mathematical Sciences, University of Cambridge,\\
Wilberforce Road, Cambridge, CB3 0WA, U.K.\vskip 5pt }
{email: {\tt P.K.Townsend@damtp.cam.ac.uk}}

\end{center}

\vskip 0.5cm

\begin{center} {\bf ABSTRACT}\\[3ex]
\end{center}

The  Souriau  massless Galilean particle of   ``colour''  $k$ and spin $s$ is shown to be the Galilean limit 
of the Souriau tachyon of mass $m = ik$ and spin $s$.  We compare and contrast this result  with the Galilean limit
of the Nambu-Goto string and  Green-Schwarz superstring.

\end{titlepage}

\newpage
\setcounter{page}{1} 
\numberwithin{equation}{section}




\section{Introduction}

The action describing a free non-relativistic point particle is Galilean invariant with the particle's mass appearing as a central charge in the Poisson bracket algebra of 
the corresponding Noether charges; this enlarged algebra is called the Bargmann algebra. A strictly Galilean dynamical system that realizes the Galilei algebra without this central charge is
a ``massless Galilean'' system; the concept and terminology are due to Souriau \cite{Sou}, who also provided a simple example, which has applications to spinoptics \cite{Duval:2005ry,Duval:2013aza}:  the massless Galilean particle of  ``colour''  $k$ and spin $s$. 

It was recently shown  that the Nambu-Goto string admits a strictly Galilean limit, and the same is true for any Dirac-type p-brane for $p>0$, so these provide further examples of massless Galilean systems \cite{Batlle:2016iel}. It has also been shown that the Green-Schwarz superstring admits a super-Galilean limit in which the Galilei algebra  is enlarged to a superalgebra \cite{Gomis:2016zur}. Although this superalgebra does have a central charge, the Galilei subalgebra does not, so the Galilean superstring provides an example of a ``massless super-Galilean''  system.

Curiously, the limiting procedure that leads to the Galilean $p$-brane does not apply for $p=0$. The Galilean massless particle is {\it not} the
Galilean $p$-brane for $p=0$; in other words, it is not a Galilean limit of the massive relativistic particle. Could it be a limit of the {\it massless} relativistic particle? In paragraph 
14.54 of the English edition of his book,  Souriau affirms that it is, but he says that the limit is ``of a different kind''  that  ``gives rise to a family of distinct non-relativistic particles, each one 
labeled by a color $k$'' \cite{Sou}. Unfortunately, Souriau does not give details, and this is also true of a statement of relevance here that he makes in the very next paragraph of his book: 
``As for tachyons, it does not seem that one can obtain a non-relativistic limit  for them''. 

In this paper we show that the massless Galilean  particle of colour $k$ is a limit of the relativistic {\it tachyon} of  imaginary mass $m=ik/c$ (where $c$ is the speed of light). 
This explains why the massless Galilean particle action cannot be obtained by choosing $p=0$ in the Galilean p-brane action of \cite{Batlle:2016iel}: the relativistic starting point
for the former  is not the $p=0$ case of the relativistic starting point of the latter.  The massless Galilean particle and the Gailean p-brane for $p>0$ are two quite different  massless 
Galilean systems.  Further differences become apparent when one considers the extension to massless super-Galilean systems, but we postpone this discussion to the end of the paper. 

We begin with a brief review of the massless Galilean particle in a notation that is convenient for our purposes, comparing and contrasting it with the Galilean string. 
We then review the massive relativistic particle with spin incorporated via the manifestly Lorentz invariant ``Souriau 2-form'', before discusing its tachyonic version and taking the
Galilean limit to recover the massless Galilean particle. We comment on the incorporation of spacetime supersymmetry in our concluding discussion. 

\section{Galilean massless particle}

The phase space of the  massless Galilean particle is parametrized by position 3-vector  ${\bf x}$ and time scalar $t$, and their conjugate momenta ${\bf p}$ and $E$, subject to one 
phase-space constraint. The phase-space action is\footnote{The constraint is that of the massless relativistic particle with $p_0=k$, which is the classical analog of the restriction 
of solutions of the wave equation to those of frequency $k$, hence the ``colour'' terminology for $k$.}
\begin{equation}\label{GMP}
S = \int d\tau\left\{ {\bf p} \cdot \dot{\bf x} - E \dot t - \frac{1}{2} e\left(|{\bf p}|^2-k^2\right) \right\} - s S_{WZ}\, , 
\end{equation}
where the overdot indicates a derivative with respect to the arbitrary worldline parameter $\tau$, and $S_{WZ}$ is what we now customarily call a Wess-Zumino action; it is derived from the phase-space 2-form (the exterior product of forms is implicit)
\begin{equation}\label{WZform}
\Omega_{WZ} = \frac{1}{2k^3} {\bf p}\cdot d{\bf p}\times d{\bf p}\, . 
\end{equation}
This 2-form is closed ($d\Omega_{WZ}=0$) as a consequence of the phase-space constraint  imposed by a Lagrange multiplier $e$:
\begin{equation}
|{\bf p}|^2=k^2\, . 
\end{equation}
Souriau did not write down this action as he preferred to work directly with the  symplectic 2-form 
\begin{equation}
\Omega= d{\bf p}\cdot d{\bf x} - dE\, dt - \frac{s}{2k^3}{\bf p}\cdot d{\bf p}\times d{\bf p} \, . 
\end{equation}
Inversion of $\Omega$ on the constraint surface yields the canonical Poisson brackets 
\begin{equation}\label{PBs}
\left\{E,t\right\}_{PB} =1 \, , \qquad \left\{x^i,p_j\right\}_{PB} = \delta^i_j \, , \qquad \left\{x^i,x^j\right\}_{PB} =  -\frac{1}{k^3} s\varepsilon^{ijk}p_k\, ,
\end{equation}
where $\{x^i,p_i; i=1,2,3\}$ are the cartesian components of ${\bf x}$ and ${\bf p}$. Notice that these relations imply, for non-zero $s$,  that the space coordinates become non-commuting operators in the quantum theory. 

The manifest Galilean invariance of $\Omega$  shows that the action (\ref{GMP}) is Galilean invariant up to a surface term, despite the fact that there is no manifestly Galilean invariant expression for  $S_{WZ}$. The corresponding Noether charges are 
\begin{equation}\label{Noether1}
H=E\, , \qquad {\bf P}= {\bf p}\, , \qquad {\bf G} = {\bf p} t \, , \qquad {\bf J} = {\bf x} \times {\bf p} + \frac{s}{k} {\bf p}\, . 
\end{equation}
A simple way to verify the expression for ${\bf J}$ is to consider the variation of  $\Omega$ induced by an infinitesimal  rotation with parameter $\bfo$ (i.e. $\delta {\bf x} = \bfo\times {\bf x}$ etc.). One finds that
\begin{equation}
\delta \Omega = d(d\bfo \cdot {\bf J})\, , 
\end{equation}
where ${\bf J}$ is as given. This confirms\footnote{The variation of the Lagrangian 1-form is a  total derivative for constant $\bfo$, and we read off the corresponding Noether charge
from the derivative of $\bfo$.} that $s$ represents spin\footnote{What we are calling spin is decomposed by Souriau  into a magnitude that he calls spin and a sign that he calls ``helicity''.}.  We remark that the same WZ term,  but with ${\bf p}$ replaced by an $SO(1,2)$ vector, was used in \cite{Schonfeld:1980kb} to incorporate spin in  the action for a {\it relativistic} particle in a 3-dimensional {\it Minkowski} spacetime. 

Using the Poisson bracket relations (\ref{PBs}), it may be verified  that the Noether charges (\ref{Noether1}) span  the Galilei algebra. In particular, one finds that 
\begin{equation}\label{massless}
\left\{P_i, G_j\right\}_{PB} =0\, , 
\end{equation}
which implies that the total momentum is boost-invariant!  Compare this state of affairs with that of the standard non-relativistic point particle: its mass $m$ appears as a central charge in this Poisson bracket relation, implying that the total momentum is not boost invariant, as one would expect. The absence of this central charge is the characteristic feature of a massless Galilean system.

\subsection{Comparison with the Galilean string}

Let us pause to make a comparison (for zero spin) with the (closed) Galilean string \cite{Batlle:2016iel}.  In this case all canonical variables are periodic functions of the string coordinate $\sigma$, and  
the Galilean Noether charges are 
\begin{equation}
H= \oint \!d\sigma E \, , \quad {\bf P} = \oint \! d\sigma \, {\bf p}\, ,   \quad {\bf G} = \oint \! d\sigma\,  {\bf p} t \, , \quad {\bf J} = \oint \! d\sigma\,  {\bf x} \times {\bf p}\, . 
\end{equation}
The phase space constraint is found from a Galilean limit of the string mass-shell constraint $p^2+ (Tx')^2=0$, where the prime indicates a derivative with respect to $\sigma$,  and this limit yields
\begin{equation}\label{string}
|{\bf p}|^2 = (Tt')^2\, . 
\end{equation}
If the string is wound $n$ times around  the ``time direction'', thus allowing the gauge choice $t'=n$, one can show that the total momentum ${\bf P}$ satisfies the bound \cite{Gomis:2016zur}
\begin{equation}\label{bound}
|{\bf P}|^2 \le n^2\, . 
\end{equation}
This has non-trivial solutions if $n\ne0$, but  there is no particle analog of this possibility.  The mass-shell constraint for  a particle of mass $m$ is $
p^2+(mc)^2=0$ (we use the ``mostly plus'' Minkowski metric signature) and the same limit yields
\begin{equation}
|{\bf p}|^2 = - (mc)^2\, , 
\end{equation}
which has no solutions  for real non-zero $m$ (and only the trivial solution ${\bf p}={\bf 0}$ for $m=0$). However, it does have solutions if we allow $m$ to be imaginary, in  which case the relativistic particle is a tachyon. 

We shall now pursue this idea for a relativistic progenitor of the massless Galilean particle in the context of a classical description,  again due to Souriau \cite{Sou}, of a {\it relativistic} particle of mass $m$ and spin $s$ in a 4-dimensional Minkowski background.

\section{The Souriau spinning particle} 

Souriau's Lorentz covariant description of the massive spinning particle requires the introduction of an independent ``polarization'' 4-vector ($w$) in addition to the particle's position 4-vector ($x$) and momentum 4-vector ($p$). These are subject to the three constraints
\begin{equation}\label{constraints2}
p^2 = -(mc)^2\, , \qquad p\cdot w =0\, , \qquad w^2 = (mc)^2 s^2\, .  
\end{equation}
Now we introduce the 2-form 
\begin{equation}\label{symplectic}
\Omega = dp^\mu dx_\mu +  \Omega_S \, , 
\end{equation}
where the second term is the  ``Souriau 2-form''
\begin{equation}
\Omega_S =  \frac{1}{2p^2} \varepsilon^{\mu\nu\rho\sigma} w_\rho p_\sigma\left( \frac{1}{p^2} dp_\mu dp_\nu + \frac{1}{w^2} dw_\mu dw_\nu\right)\, . 
\end{equation}
This 2-form  is closed (and hence so is $\Omega$) as a consequence of the constraints, as follows from the following lemma: 
\begin{itemize}
\item {\bf Lemma}: Given two 4-vectors $(u,v)$, the 2-form 
\begin{equation}
\omega = \frac{1}{2}\varepsilon^{\mu\nu\rho\sigma} v_\rho u_\sigma \left(du_\mu du_\nu - dv_\mu dv_\nu\right)
\end{equation}
is closed  if 
\begin{equation}
u^2=-1\, , \qquad v^2 =1\, , \qquad u\cdot v =0\, . 
\end{equation}
To prove this lemma, we first observe that these constraints imply 
\begin{equation}\label{derived}
u\cdot du=0\, , \qquad v\cdot dv=0\, , \qquad u\cdot dv + v\cdot du =0\, .
\end{equation} 
We now choose a Lorentz frame for which 
\begin{equation}\label{Lframe1}
u_\mu\big| = (1;0,0,0) \, , \qquad v_\mu\big| = (0; 0,0,1)\, , 
\end{equation}
where the $\big|$ notation indicates that this choice is made at one point; i.e. it is not assumed to hold for $du$ and $dv$. However, the
derived constraints (\ref{derived}) in this frame are 
\begin{equation}\label{Lframe2}
du_0=0\, , \qquad dv_3 =0\, , \qquad dv_0 = du_3\, .
\end{equation}
Using both  (\ref{Lframe1}) and (\ref{Lframe2}), a straightforward calculation yields $d\omega=0$.
\end{itemize} 
By observing that $\Omega_S =s\omega$ for $(p,w)=mc(u,sv)$, we conclude that $d\Omega_S=0$. 

It is important to appreciate that  $\Omega$   is not a ``symplectic''  2-form for the 12-dimensional space parametrized by the components of the  three 4-vectors $(x,p,w)$. This is because it is not invertible on this space;  it is block diagonal in the basis $\{dx,dp,dw\}$ but the $4\times4$ ($dw,dw$) block  has  $p$ and $w$ as two zero-eigenvalue  eigenvectors. However, within the 4-dimensional Minkowski subspace of fixed $(x,p)$ the two $w$-dependent constraints determine a ($p$-dependent) 2-sphere whose tangent vectors are  orthogonal to both $p$ and $w$. To see this it suffices to  choose the frame for which  ${\bf p}={\bf 0}$; then $w=(0, {\bf w})$ with $|{\bf w}|^2= (mcs)^2$.  The pull-back of $\Omega$ to the 10-dimensional submanifold of topology $\bR^8\times S^2$ defined by the $w$-dependent constraints {\it is} invertible. Its inversion yields a set of canonical Poisson brackets for this phase space, with respect to which the remaining $w$-independent constraint is first-class, so the physical phase-space is 8-dimensional; in fact  it is topologically $\bR^6\times S^2$, where the first factor is the phase space for a free particle in the Euclidean 3-space and the second factor is the spin phase space (as becomes manifest in a bi-twistor formulation \cite{Tod:1977vf,Fedoruk:2014vqa}).

Finally, to see why the parameter $s$ is the particle's spin, we observe that the infinitesimal Lorentz transformations
\begin{equation}
\delta x^\mu = \Lambda^\mu{}_\nu x^\nu \, , \qquad \delta p_\mu = \Lambda_\mu{}^\nu p_\nu \, , \qquad \delta w^\mu = \Lambda^\mu{}_\nu w^\nu\, , 
\end{equation}
induce the following variation of $\Omega$: 
\begin{equation}
\delta \Omega = -d \left[ \frac{1}{2} d\Lambda_{\mu\nu} J^{\mu\nu}\right]\, , 
\end{equation}
where
\begin{equation}
J^{\mu\nu} = 2 x^{[\mu} p^{\nu]}  - \frac{1}{(mc)^2} \varepsilon^{\mu\nu\rho\sigma} p_\rho w_\sigma\, . 
\end{equation}
If we use this (and $P_\mu=p_\mu$) to compute the Pauli-Lubanski pseudo-vector $L$ we find that 
\begin{equation}
L^\mu := \frac{1}{2} \varepsilon^{\mu\nu\rho\sigma} P_\nu J_{\rho\sigma} = w^\mu\, , 
\end{equation}
and hence that 
\begin{equation}
L^2 = w^2 = (mc)^2 s^2\, . 
\end{equation}

\subsection{The tachyonic spinning particle and its Galilean limit}

Now we consider the tachyonic version of Souriau's relativistic spinning particle model obtained by setting $mc = ik$ for some real number $k$. This yields the phase-space constraints 
\begin{equation}\label{tachcon}
p^2 = k^2\, , \qquad w^2 = -(ks)^2\, , \qquad p\cdot w =0\, .  
\end{equation}
Now $p$ is spacelike and $w$ is timelike, but the Souriau 2-form is still closed, by an application of the above lemma but with a reversed identification of $(u,v)$ with multiples of $(p,w)$. 

It is again true that $\Omega$ is not invertible on the 12-dimensional space parametrized by the components of the three 4-vectors $(x,p,w)$ but {\it is} invertible on the 10-dimensional 
submanifold determined  by the $w$-dependent constraints. However, the surface that these constraints define within the Minkowski subspace of fixed $(x,p)$ is now a
hyperboloid rather than a sphere. To see this we may choose a frame for which $p \propto (0,{\bf n})$ for unit 3-vector ${\bf n}$; then $w$ is a timelike vector of fixed interval in the 3D Minkowski subspace orthogonal  to ${\bf n}$ and hence lies on a 2-dimensional hyperboloid. 

To take the Galilean limit of the spinning tachyon, we first rescale $x^0, p_0$ and ${\bf w}$ as follows
\begin{equation}\label{rescale} 
x^0 \to \lambda x^0\, , \qquad  p_0 \to p_0 /\lambda \, , \qquad {\bf w} \to {\bf w}/\lambda\, , 
\end{equation}
where $\lambda$ is positive, and then we take $\lambda\to\infty$. As $x^0=ct$ and $p_0=E/c$ (for dimensionless Minkowski metric) this is equivalent to the $c\to\infty$ limit but with an additional specification of how to take this limit for the components of $w$. One finds that  the constraints (\ref{tachcon}) reduce to
\begin{equation}
|{\bf p}|^2 =k^2 \, , \qquad w_0^2 = (ks)^2\, ,\qquad p_0w_0 = {\bf p}\cdot {\bf w} \, . 
\end{equation}
Assuming, for simplicity, that both $w_0$ and $ks$ are positive, the second of these constraints tells us that $w_0=ks$. The third constraint can be solved for the component of ${\bf w}$ parallel to ${\bf p}$, 
but this leaves two components of ${\bf w}$ undetermined. This is as expected because the $w$-dependent constraints initially restricted $w$ to a 2-dimensional hyperboloid.  However, when 
we perform the rescaling (\ref{rescale}) in the action, and take the $\lambda\to\infty$ limit, these unrestricted variables drop out. If we use $w_0=ks$ to eliminate $w_0$, the  
Souriau 2-form reduces to the 2-form $\Omega_{WZ}$ of (\ref{WZform}), and the net result is that we recover the action (\ref{GMP}) for the Galilean massless particle of colour $k$ and spin $s$.

A peculiar feature of this limit is that the physical phase space is only 6-dimensional in the limit whereas it was 8-dimensional initially. We suspect that this was the source of Souriau's reservations
about the non-relativistic limit of a spinning tachyon.

\section{Discussion}

We have shown that the massless Galilean particle of colour $k$ is a non-relativistic limit of  a tachyon of mass $m=ik/c$.  Although tachyons are usually considered to be unphysical, there are {\it unitary} irreducible tachyonic representations of the Poincar\'e group \cite{Wigner:1939cj} and the possibility that these may have some  physical realization has been explored in many papers; see e.g. \cite{Perepelitsa:2014pva} for a recent review with references to the literature.  Consequently, one cannot conclude from its tachyonic origin that the massless Galilean particle is intrinsically unphysical. 

However, this conclusion changes when we consider the supersymmetric extension of massless Galilean systems because there are no unitary irreducible tachyonic representations of the super-Poincar\'e group. The tachyonic superparticle is intrinsically non-unitary, and we should therefore expect the same of any attempt at a supersymmetrization of the massless Galilean particle.  This argument does not apply to the Galilean superstring, for which unitarity simply requires the same bound (\ref{bound}) on the total momentum  that is already implied by the classical  phase-space constraints \cite{Gomis:2016zur}. Inspection of the details shows that this is due to the intrinsically ``stringy'' topological charge in the super-Galilean algebra of Noether charges. 

As a final comment, inspired by the idea expounded in \cite{Duval:2014uoa} of a ``duality'' relating the Galilean to the Caroll limit \cite{caroll},  we observe that the status of a tachyon in the Galilean limit is analogous to that of a bradyon in the Caroll limit \cite{caroll}. In the former case, the superluminal velocity is forced to go to infinity.  In the latter case the subluminal velocity must go to zero. In both cases, the mass (real or imaginary) corresponds to a property of the resulting system. 

\subsection*{Note added:}

Another massles Galilean system, in the Souriau sense, is provided by the particle with Lagrangian 
\begin{equation}
L= \frac{\mu}{2} |\ddot{\bf x}|^2\, .  
\end{equation}
Because this Lagrangian is {\it strictly} Galilean invariant (its variation is {\it not} a total time derivative) there is no central charge in the  algebra of Galilei Noether charges \cite{Gauntlett:1990nk}; this also follows from dimensional analysis because the only parameter, $\mu$,  has dimensions of mass $\times$ time-squared rather than mass. The corresponding phase-space Lagrangian is
\begin{equation}
L= {\bf p} \cdot\dot {\bf x} + {\bf q}\cdot \dot{\bf y} - H\, , \qquad H= {\bf p} \cdot {\bf y} + \frac{1}{2\mu} |{\bf q}|^2\, , 
\end{equation}
where we use a rescaled version of the phase-space coordinates of  \cite{Gauntlett:1990nk}.  By taking the $\mu\to\infty$ limit 
we get the phase-space Lagrangian
\begin{equation}\label{SZ}
L_{SZ} = {\bf p} \cdot \left(\dot {\bf x}-{\bf y}\right)  + {\bf q} \cdot \dot {\bf y}\, ,  
\end{equation}
which was the basis for a dynamical alternative to dark energy proposed by Stichel and Zakrzewski \cite{Stichel:2009sz}. These authors also considered a relativistic analog, which they interpreted as a tachyon.  In order to elucidate the relation of this result to the results reported here, we present a brief analysis  of the Stichel-Zakrzewski Lagrangian. 

The equations of motion for  ${\bf y}$ and ${\bf p}$ are jointly equivalent to 
\begin{equation}
{\bf y}= \dot{\bf x}\, , \qquad {\bf p} = \dot{\bf q}\, , 
\end{equation}
so we may consistently eliminate these variables to get an equivalent Lagrangian for ${\bf x}$ and ${\bf q}$ alone.  In terms of the linear combinations 
\begin{equation}
{\bf z}_\pm = \left({\bf x} \mp \frac{1}{2m}{\bf q}\right) \, , 
\end{equation}
where $m$ is an arbitrary non-zero constant mass parameter, this equivalent Lagrangian is 
\begin{equation}
L_{NR} = \frac{m}{2} \left[|\dot {\bf z}_+|^2 - |\dot {\bf z}_-|^2\right] + \frac{d}{dt}\left(\cdots\right) \, . 
\end{equation}
Each term is separately Galilean invariant, with Noether charges ${\{\bf P}_\pm,{\bf G}_\pm, {\bf J}_\pm\}$ and central charges $\pm m$. The linear combinations
\begin{equation}
{\bf P} = {\bf P}_+ + {\bf P}_- \, , \qquad {\bf G} = {\bf G}_+ + {\bf G}_- \, , \qquad {\bf J} = {\bf J}_+ + {\bf J}_-\, , 
\end{equation}
span a Galilei algebra with zero central charge because the total central charge is $m-m=0$. So we indeed have a massless Galilean system, but at the cost 
of a non-unitary quantum theory. 

The relativistic analog of $L_{SZ}$ considered in  \cite{Stichel:2009sz} was presented as a set of equations to be satisfied by phase-space variables that  were assumed to be  
functions of an arbitrary worldline time parameter, although the constraint generating  time reparametrizations was not given.  We can proceed more systematically now that we have
established the equivalence of $L_{SZ}$ to $L_{NR}$; the latter is obviously the Galilean limit of the relativistic  mechanics model 
with Lorentz invariant Lagrangian
\begin{equation}\label{Lrel}
L_{\rm Rel} = -mc^2 \left[ \sqrt{1- |{\bf u}_+|^2} - \sqrt{1- |{\bf u}_-|^2}\right]\, , \qquad \left({\bf u}_\pm =\dot{\bf z}_\pm/c\right).
\end{equation}
Each term is separately Lorentz invariant (although not manifestly so because the transformations are non-linear) and the $c\to\infty$ limit yields 
$L_{NR}$ directly because the rest-mass energy cancels between the two terms. We also have two sets of Lorentz generators, in particular
two conserved 4-momenta $P_\pm$ and the Lorentz algebra with the Galilean limit is found by taking the sum.  In particular, the total 4-momentum 
$P= P_++P_-$ is the combination relevant to the Galilean limit, and 
\begin{equation}
P^2= 2m^2c^2 \left[\gamma_+\gamma_- \left(1-{\bf u}_+\cdot {\bf u}_-\right) -1\right] \ge 0 ,  \qquad \left(\gamma_\pm = \frac{1}{\sqrt{1- |{\bf u}_\pm|^2}}\right)\, . 
\end{equation}
It follows that $P$ is spacelike unless ${\bf u}_+={\bf u}_-$, in which case it is null.  This is the tachyonic behaviour found in \cite{Stichel:2009sz} although we would choose to interpret the  model as a two-particle system rather than a tachyon. In any case, the relative minus sign between the two terms of the Lagrangian $L_{\rm Rel}$ gives us information that is not obtainable from the equations of motion alone, and  it  tells us that the quantum theory is not unitary,  as was to be expected from  its Galilean limit.

\section*{Acknowledgements} 

We thank Peter Horvathy and Peter Stichel for bringing aspects of their work to our attention.  LM is grateful for the hospitality of  the Department of Applied Mathematics and Theoretical Physics at the University of Cambridge and the Departament de F\'{\i}sica Qu\`antica i Astrof\'{\i}sica at the University of Barcelona.  CB is partially supported by  the Generalitat de Catalunya through project 2014 SGR 267 and by the Spanish government (MINECO/FEDER) under project CICYT DPI2015-69286-C3-2-R. JG has been supported  in part by FPA2013-46570-C2-1-P, 2014-SGR-104 (Generalitat de Catalunya) and Consolider CPAN and by the Spanish goverment (MINECO/FEDER) under project MDM-2014-0369 of ICCUB (Unidad de Excelencia Mar\'\i a de Maeztu). 
PKT acknowledges support from the UK Science and Technology Facilities Council (grant ST/L000385/1).

\providecommand{\href}[2]{#2}\begingroup\raggedright\endgroup

\end{document}